\documentclass[fleqn,12pt,twoside]{article}
\usepackage{epsf,espcrc1}
\usepackage{graphicx}
\usepackage[figuresright]{rotating}

\newcommand{\AmS}{{\protect\the\textfont2
  A\kern-.1667em\lower.5ex\hbox{M}\kern-.125emS}}

\title{$\Phi$ meson production at RHIC}

\author{
L.~Bravina$^a$, L.~Csernai$^b$, 
A.~Faessler$^c$, C.~Fuchs$^c$, S.~Panitkin$^d$, Nu Xu$^e$,
E.~Zabrodin$^{a,c}$
\\
\vspace*{.25 cm}
{\small\it
$^a$Institute for Nuclear Physics, Moscow State University,
RU-119899 Moscow, Russia} \\
{\small\it
$^b$Department of Physics, University of Bergen, 
N-5007 Bergen, Norway}\\
{\small\it
$^c$Institute for Theoretical Physics, University of
T\"ubingen, D-72076 T\"ubingen, Germany}\\
{\small\it
$^d$Brookhaven National Laboratory, Upton, NY 11973, USA}\\
{\small\it
$^e$Nuclear Science Division, Lawrence Berkeley National Laboratory,
CA 94720, USA}
}
       
\begin{document}

\maketitle

\begin{abstract}
{\small
The production of $\phi$ mesons in Au+Au collisions at RHIC and their
propagation in a hot and dense nuclear medium is studied within the 
microscopic quark-gluon string model. The inverse slope parameter of 
the transverse mass distribution agrees well with that extracted from 
the STAR data, while the absolute yield of $\phi$ is underestimated 
by a factor 2. It appears that the fusion of strings alone cannot 
increase the $\phi$ yield either. Less than 30\% of detectable 
$\phi$'s experience elastic scattering, this rate is insufficient for 
the full thermalization of $\phi$. The directed flow of $\phi$ at $|y| 
\leq 2$ demonstrates strong antiflow behavior, whereas its elliptic 
flow rises up to about 3.5\% in the same rapidity interval. As a 
function of transverse momentum it rises linearly with increasing 
$p_t$, in agreement with the STAR data, and saturates at 
$p_t > 2$\, GeV/c.
}
\end{abstract}

\vspace{.5 cm}


The investigation of nuclear matter under extreme temperatures and
densities, and the search for a predicted transition to a deconfined 
phase of quarks and gluons, the so-called Quark-Gluon Plasma (QGP), 
is one of the main goals of heavy ion experiments at ultrarelativistic 
energies. Both theorists and experimentalists are looking for genuine 
QGP fingerprints, that cannot be masked or washed out by processes on 
a hadronic level. The $\phi$ meson was proposed about two decades ago 
\cite{KMR86,Sh85} as one of the most promising QGP messengers because 
of the following reasons: Firstly, an enhancement of $s \bar{s}$ pairs 
in the QGP phase should lead to an enhancement of $\phi$ mesons, as 
well as other strange hadrons, while the production of $\phi$ in $hh$ 
interactions is suppressed due to the OZI rule \cite{Sh85}. Secondly, 
since the $\phi$ interaction cross section with non-strange hadrons is 
small, $\phi$ will keep information about the early hot and dense 
phase. Thirdly, the $\phi$ meson spectrum is not distorted by feeddown 
from resonance decays. Finally, among other channels $\phi$ decays 
into a pair of kaons and, more rarely, into a lepton pair. Both 
channels have been explored for the reconstruction of the $\phi$ yield 
in Pb+Pb collisions at SPS energy. It was found that the $\phi$ 
multiplicity extracted from the di-kaon data \cite{na49phi} is 
significantly smaller than that obtained from the dimuons 
\cite{na50phi} (the so-called $\phi$ meson puzzle). A similar effect 
has been observed recently by the PHENIX Collaboration \cite{phenix} 
in Au+Au collisions at RHIC. A possible solution of this puzzle is the 
scattering of at least one of the daughter kaons in the nuclear medium 
\cite{JJD01,FK01,Soff01,PKL02} accompanied by in-medium modifications 
of kaon and $\phi$ masses. Note also that the width and mass of the 
$\phi$ meson are quite sensitive to the strangeness content of the 
surrounding medium.

\begin{figure}[htb]
\begin{minipage}[t]{60mm}
\vspace*{-0.3cm}
\centerline{\epsfysize=66mm\epsfbox{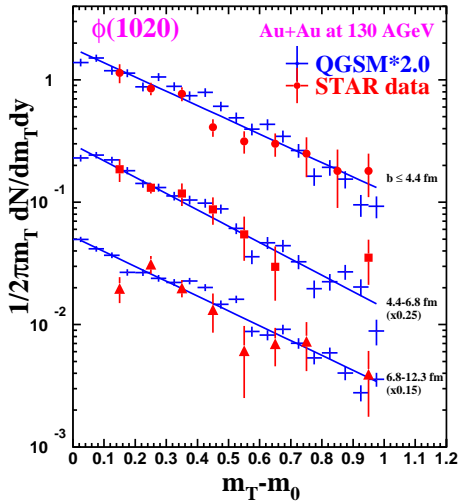}}
\vspace*{-0.9cm}
\caption{\small
$m_t$-spectra of $\phi$ mesons in three centrality 
bins in Au+Au collisions at RHIC. 
 }
\label{fig1}
\end{minipage}
\hspace{\fill}
\begin{minipage}[t]{83mm}
\vspace*{-0.88cm}
Collaboration \cite{star_phi}. Though the QGSM underestimates the
$\phi$ yield by a factor 2, it correctly reproduces the slopes of 
all three distributions. Results of the fit to an exponential 
$\displaystyle \frac{dN/dy}{2 \pi T (m_\phi + T)}\, \exp{\left( - 
\frac{m_t -m_\phi} {T}\right)}$ are plotted in Fig.~\ref{fig1} as 
well. Obtained values of the inverse slope parameter $T$ are 378 
(379$\pm$50) MeV, 349 (369$\pm$73) MeV, 370 (417$\pm$75) MeV for 
model (experimental) results, respectively. The absence of a 
centrality dependence in the $m_t$-spectra of $\phi$ may indicate 
the minor role of the $\phi$ final state interactions. The problem 
of a too low strangeness yield in the model can be solved by 
implementing the string fusion mechanism \cite{str_fus}. To check 
this, the yields of $\phi$, $h^-$, and their ratio $\phi/h^-$ have 
been calculated in the string fusion model (SFM) \cite{sfm}. Results 
\end{minipage}
\end{figure}

In the present paper we study production, propagation, and freeze-out
of $\phi$ mesons in Au+Au collisions at $\sqrt{s} = 130$\, AGeV 
within the microscopic quark-gluon string model (QGSM) \cite{qgsm}.
The transverse mass spectra of $\phi$ calculated for three different
centrality bins are presented in Fig.~\ref{fig1} together with the
experimental\, data\, reported\, by the STAR 

\begin{table}[htb]
\begin{minipage}[t]{80mm}
\vspace*{-1.6cm}
\caption{\small
Yields of $\phi$, $h^-$, and ratio $\phi/h^-$ at $|y| \leq 0.5$ in
Au+Au at RHIC calculated within the SFM. 
}
\label{tab1}
\begin{tabular}{@{}ccccc}
\hline
 & \multicolumn{4}{c}{ Centrality \ \ 0 - 11\%} \\
\hline
 &  NF & F &  F+R &  exp. \\
\hline
$\phi$      & 5.71 & 3.60& 4.54 & 5.73$\pm$0.37 \\
$h^- $      & 462  & 326 & 324   &  273$\pm$21  \\
$\phi/h^-$  & 0.0123 & 0.0110 & 0.0140 &
0.021$\pm$0.001 \\ \hline
\vspace*{-0.7cm}
\end{tabular}
\end{minipage}
\hspace{\fill}
\begin{minipage}[t]{65mm}
\vspace*{-1.90cm}
are listed in Table~\ref{tab1}. Compared to the non-fusion case 
(NF), the string fusion (F) mode significantly decreases the $\phi$ 
and $h^-$ multiplicities at midrapidity. An increase of the $\phi$ 
yield can be obtained by further rescattering (F$+$R) of 
secondary hadrons via the channels $K Y \rightarrow \phi N,\ K \Xi 
\rightarrow \phi Y,\ K \Omega \rightarrow \phi Y$, but it is still 
insufficient to match the 
\end{minipage}
\end{table}
\vspace{-0.90cm}
\hspace{-0.55cm}
data. The agreement between the calculations and the data
can be improved by increasing the string tension or dropping the
constituent quark masses \cite{Soff01}, by including the 
OZI-violation mechanism in the model (at low $\sqrt{s}$ this is the
only channel for $\phi$ production), and by increasing the number
of $s\bar{s}$-pairs inside nucleons in multipomeron diagrams
\cite{CK02}.

\begin{figure}[htb]
\begin{minipage}[t]{75mm}
\vspace*{-2.0cm}
\centerline{\epsfysize=80mm\epsfbox{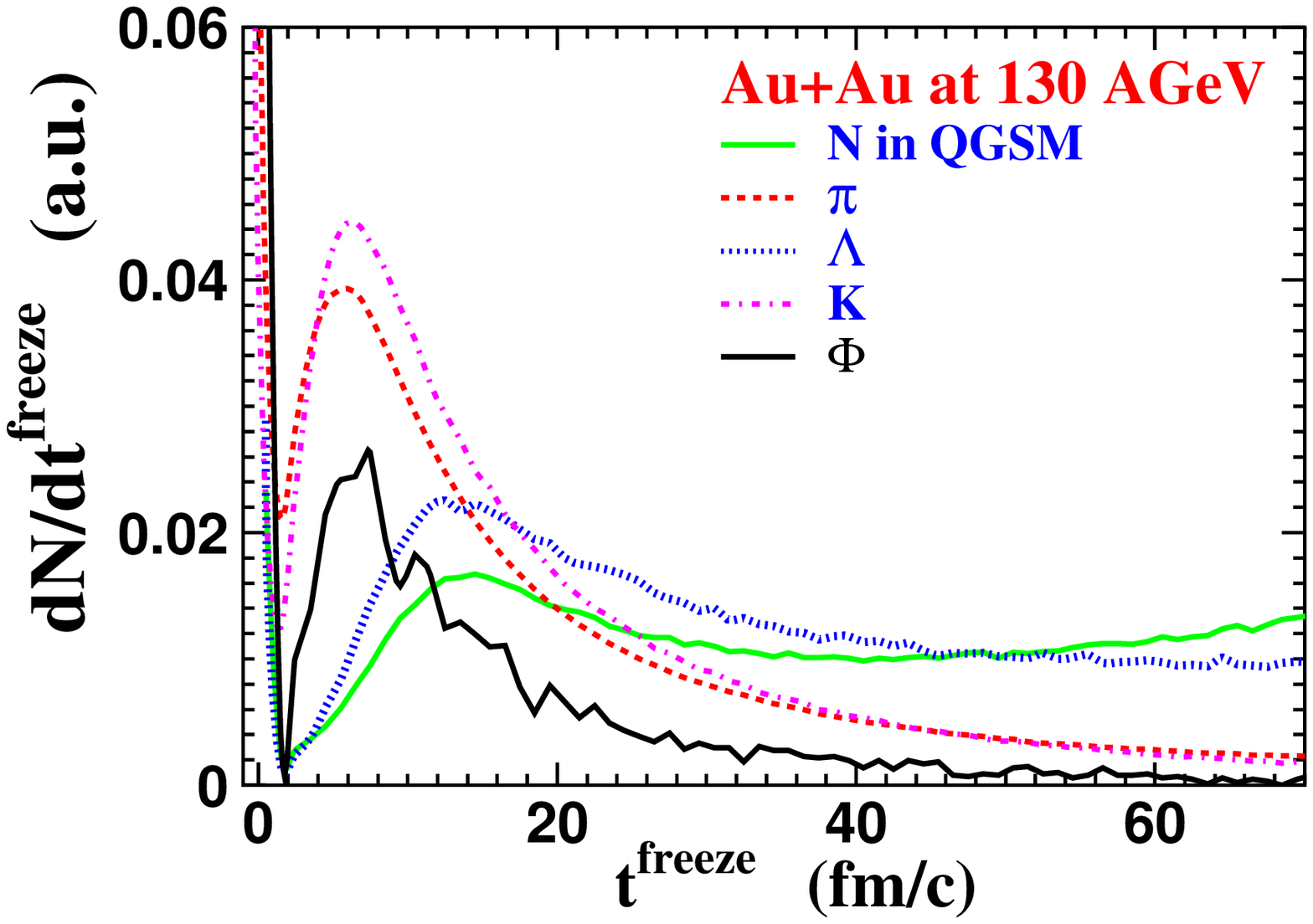}}
\vspace*{-2.0cm}
\caption{\small
$dN/dt$ distribution of $N, \pi, K, \Lambda$ and $\phi$ over their
last collision time $t^{freeze}$.
 }
\label{fig2}
\end{minipage}
\hspace{\fill}
\begin{minipage}[t]{75mm}

\vspace*{-2.00cm}
\centerline{\epsfysize=80mm\epsfbox{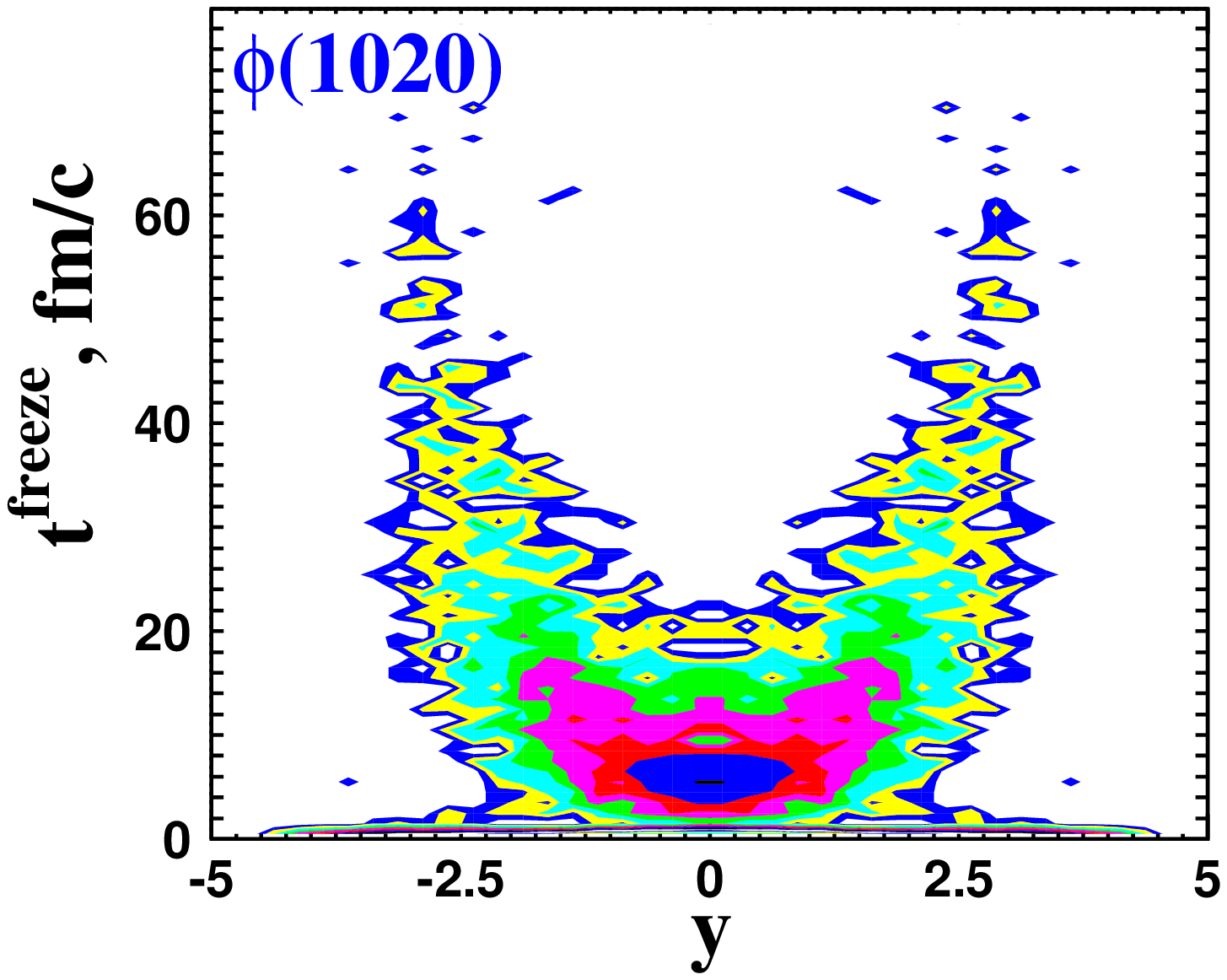}}
\vspace*{-2.00cm}
\caption{\small
$d^2N/dy dt$ distribution of the final-state $\phi$ mesons in the
$(y,t)$-plane.
 }
\label{fig3}
\end{minipage}
\vspace*{-0.4cm}
\end{figure}

The $\phi$ mesons are produced either in primary nucleon-nucleon
interactions via string fragmentation processes or in subsequent
rescattering between produced strange hadrons. We found that in the 
first 4 fm/$c$ the $\phi$ mesons are produced solely in string 
processes. The string channel dominates over the rescattering one up 
to 20 fm/$c$; in the later stages the rescattering channel prevails. 
This dynamical picture agrees well with that obtained for the $\phi$ 
production at SPS energy within the UrQMD model \cite{Soff01}. The 
$dN/dt$ distributions of $\phi, N, \pi, K, \Lambda$, which are 
decoupled from the fireball after the last elastic or inelastic 
collision, are shown in Fig.~\ref{fig2}. Compared to AGS and SPS 
energies \cite{fr_ags_sps}, a substantial part of hadrons leave the 
system immediately after their production. Mesonic distributions 
have sharp peaks at $t = 8-10$\,fm/$c$, while the distributions of 
baryons are wider due to the large number of rescatterings, which 
shift their $dN/dt$ maxima to later times. Figure \ref{fig3} depicts 
the freeze-out distribution of the $\phi$ mesons over rapidity and 
emission time. We see that the bulk production of $\phi$ occurs in 
the central rapidity window $|y| \leq 1.5$ within the first 10 
fm/$c$, while in $|y| \geq 2$ areas the $\phi$ mesons are produced 
until the
late stages of the system evolution: Because 
of the rapid expansion the central zone becomes quickly dilute, its 
energy density drops, and early freeze-out of $\phi$ mesons, which 
have small interaction cross section with non-strange matter, takes 
place. According to QGSM, about 51\% of detectable $\phi$ mesons 
decay after decoupling from the fireball.  Less than 30\% of $\phi$ 
mesons experience at least one elastic rescattering before the 
freeze-out. This relatively low rate of elastic collisions may not 
be sufficient for the thermalization of $\phi$ mesons, however, 
would it be enough for the development of noticeable anisotropic 
flow? - Other processes contributed to the flow evolution are 
production and absorption/decay of $\phi$ in dense medium. While the 
elastic collisions and resonance production of $\phi$ mesons increase 
the flow in the normal bounce-off direction, the $\phi$ absorption 
channel effectively reduces this normal component thus elongating the 
flow in the antiflow direction.

Figure \ref{fig4} shows the rapidity 
dependence of the directed flow of $\phi, N$, and $K$ in minimum bias 
Au+Au collisions at RHIC. The slopes of the all distributions are 
negative at $|y| \leq 2$, i.e. the antiflow component of the $v_1$ 
dominates over its normal counterpart (see \cite{dir_flow}). Similar
antiflow slopes of $v_1(y)$ are developed by $\pi$ and $\Lambda$ 
\cite{dir_flow,kflow}; its origin is traced to nuclear shadowing. At 
midrapidity $|y| \leq 0.5$ the directed flow of $\phi$ is quite weak.

The elliptic flow of $\phi$ mesons as a function of $\eta$ and $p_t$ 
is shown in Fig.~\ref{fig5} together with the model predictions for 
$\pi$ and $N$ \cite{ell_flow}, and recent experimental data 
\cite{yam_phi}. The magnitude 
of the $\phi$ flow at midrapidity $v_2^\phi(\eta) = 3.3 \pm 
1.1\%$ is similar to that of pions and nucleons.
At larger rapidity $v_2^\phi(y)$ is weaker than the 
flow of other hadrons. As a function of $p_t$ the elliptic flow of 
$\phi$ rises up to 2 GeV/$c$ almost linearly with 
$p_t$. The agreement with the experimental results is good, although 
large error bars do not allow more definite conclusions. The 
saturation and/or possible drop of the $v_2^\phi(p_t)$ 
at higher $p_t$ is linked to the lack of rescattering in spatially 
anisotropic finite nuclear matter.

\begin{figure}[htb]
\begin{minipage}[t]{75mm}
\vspace*{-0.1cm}
\centerline{\epsfysize=47mm\epsfbox{phi_v1_new.epsi}}
\vspace*{-0.9cm}
\caption{\small
Directed flow $v_1(y, all\ p_t)$ of $\phi, N, K$ in minimum bias 
Au+Au events at $\sqrt{s} = 130$ AGeV.
 }
\label{fig4}
\end{minipage}
\hspace{\fill}
\begin{minipage}[t]{74mm}
\vspace*{-0.17cm}
\centerline{\epsfysize=48mm\epsfbox{phipteta.epsi}}
\vspace*{-0.9cm}
\caption{\small
The same as Fig.~\protect\ref{fig4} but for the elliptic flow 
$v_2(y, all\ p_t)$ and $v_2(p_t, |y| \leq 1)$ of $\phi, N, K$. 
Data are taken from \protect\cite{yam_phi}.
 }
\label{fig5}
\end{minipage}
\vspace{-0.5cm}
\end{figure}

In summary, the QGSM successfully describes $m_t$-spectra of $\phi$
in different centrality bins in Au+Au collisions at $\sqrt{s} = 
130$\,AGeV, but underestimates the total $\phi$ production. This
problem cannot be cured alone by the fusion of strings. Mesons are
frozen in the model earlier than baryons. The QGSM predicts unique
antiflow slope at midrapidity for the directed flow of $\phi, \pi,
N, K, \Lambda$, and reproduces correctly the elliptic flow 
$v_2^\phi (p_t)$. 

{\small {\it Acknowledgments.}
Fruitful discussions with A. Capella, A. Kaidalov, M. 
Krivoruchenko, D. Sousa and S. Voloshin are gratefully acknowledged.
This work was supported by the EGK
Basel-T\"ubingen under DFG contract GRK683, by the BMBF under 
contract 06T\"U986, and by the Bergen Computational
Physics Laboratory (BCPL) in the framework of the
European Community - Access to Research Infrastructure action of the
Improving Human Potential Programme.
}

\end{document}